# Diagnosis of Pulmonary Hypertension by Integrating Multimodal Data with a Hybrid Graph Convolutional and Transformer Network

Fubao Zhu [a], Yang Zhang [a], Gengmin Liang [b], Jiaofen Nan [a], Yanting Li [a], Chuang Han [a], Danyang Sun [a], Zhiguo Wang [a], Chen Zhao [c], Wenxuan Zhou [d], Jian He [e], Yi Xu [e], Iokfai Cheang [b], Xu Zhu [b], Yanli Zhou [b,*], Weihua Zhou [f,g]

[a] School of Computer Science and Technology, Zhengzhou University of Light Industry, Zhengzhou 450001, Henan, China
[b] State Key Laboratory for Innovation and Transformation of Luobing Theory, Department of Cardiology, the First Affiliated Hospital of Nanjing Medical University, Nanjing 210029, Jiangsu, China
[c] Department of Computer Science Kennesaw State University Marietta, GA, USA
[d] Department of Integrated Traditional Chinese and Western Clinical Medicine, Hebei Medical University, Hebei, China
[e] Department of Radiology, The First Affiliated Hospital of Nanjing Medical University, Nanjing, 210029, China
[f] Department of Applied Computing, Michigan Technological University, Houghton, MI, USA
[g] Center for Biocomputing and Digital Health, Institute of Computing and Cybersystems, and Health Research Institute, Michigan Technological University, Houghton, MI, USA

*Correspondence:
Yanli Zhou
zhyl88@qq.com
State Key Laboratory for Innovation and Transformation of Luobing Theory, Department of Cardiology, the First Affiliated Hospital of Nanjing Medical University, Nanjing 210029, Jiangsu, China

**Abstract**

Early and accurate diagnosis of pulmonary hypertension (PH) is essential for optimal patient management. Differentiating between pre-capillary and post-capillary PH is critical for guiding treatment decisions. This study develops and validates a deep learning-based diagnostic model for PH, designed to classify patients as non-PH, pre-capillary PH, or post-capillary PH.

This retrospective study analyzed data from 204 patients (112 with pre-capillary PH, 32 with post-capillary PH, and 60 non-PH controls) at the First Affiliated Hospital of Nanjing Medical University. Diagnoses were confirmed through right heart catheterization. We selected 6 samples from each category for the test set (18 samples, 10%), with the remaining 186 samples used for the training set. This process was repeated 35 times for testing.

This paper proposes a deep learning model that combines Graph convolutional networks (GCN), Convolutional neural networks (CNN), and Transformers. The model was developed to process multimodal data, including short-axis (SAX) sequences, four-chamber (4CH) sequences, and clinical parameters.

Our model achieved a performance of Area under the receiver operating characteristic curve (AUC) = 0.81±0.06(standard deviation) and Accuracy (ACC) = 0.73±0.06 on the test set. The discriminative abilities were as follows: non-PH subjects (AUC = 0.74±0.11), pre-capillary PH (AUC = 0.86±0.06), and post-capillary PH (AUC = 0.83±0.10). It has the potential to support clinical decision-making by effectively integrating multimodal data to assist physicians in making accurate and timely diagnoses.

**Keywords**

Pulmonary hypertension; Multimodality; Cardiac magnetic resonance imaging; Deep learning; Graph convolutional network

**Abbreviations**

| | |
|---|---|
| 4CH | four-chamber |
| ACC | Accuracy |
| AI | Artificial intelligence |
| AUC | Area under the receiver operating characteristic curve |
| BN | Batch normalization |
| CMR | Cardiac magnetic resonance imaging |
| COPD | Chronic obstructive pulmonary disease |
| DCM | Dilated cardiomyopathy |
| GCN | Graph convolution neural |
| IHD | Ischemic heart disease |
| PH | Pulmonary hypertension |
| RAC | Relative area change of pulmonary artery |
| ReLU | Rectified Linear Unit |
| RHC | Right heart catheterization |
| RID | Rheumatoid immune disease |
| ROI | Region of interest |
| SAX | short-axis |
| VHD | Valvular heart disease |

## 1. Introduction

Pulmonary hypertension (PH) is a significant global health concern, affecting individuals across all age groups [1]. Current estimates indicate that PH affects approximately 1% of the global population [2]. Accurate differentiation between pre-capillary and post-capillary PH is essential for developing effective treatment strategies. PH due to left heart disease—characterized by post-capillary hypertension—is the most common type, but its diagnosis remains challenging in some cases [3 4].

Right heart catheterization (RHC) is the gold standard for PH diagnosis; however, its invasive nature and limited availability in some hospitals restrict its routine application in clinical practice [5]. Doppler echocardiography-based estimates of pulmonary artery systolic pressure are less accurate in PH patients compared to RHC [6]. CMR has become a valuable non-invasive tool for PH assessment, offering detailed insights into pulmonary artery pressure, right heart function, and cardiac structure [7-9]. However, CMR interpretation relies heavily on physician expertise, which can result in subjective variability and inconsistencies among clinicians [10 11]. These limitations hinder the widespread and standardized adoption of CMR in PH diagnosis [12]. Recent advancements in artificial intelligence (AI) have demonstrated its potential in medical diagnostics, including the prediction of PH and left atrial (LA) pressure using electrocardiograms (ECG) and chest X-rays [13-16]. Several studies have reported the effective use of machine learning (ML) in assessing diastolic function [17-19]. For example, Hirata et al. [20] developed a machine learning model that predicts non-PH subjects, pre-capillary pulmonary hypertension, and post-capillary pulmonary hypertension, with AUC values of 0.789, 0.766, and 0.742, respectively. Machine learning methods have some limitations, such as complicated data preprocessing, large workload for feature selection and extraction. In contrast, deep learning methods offer stronger feature extraction and model generalization capabilities, enabling automatic learning of complex data features.

Existing deep learning methods for diagnosing PH, such as the work by Kusunose et al. [21] have achieved an AUC of 0.71 using X-ray data to build a predictive model that distinguishes PH from non-PH subjects. However, these methods are limited to a single data source and cannot differentiate between pre-capillary and post-capillary PH patients. This study aims to develop and validate a deep learning model based on multimodal data to classify patients into three groups: non-PH, pre-capillary pulmonary hypertension, and post-capillary pulmonary hypertension.

## 2. Materials

### 2.1. Study population and data collection

This retrospective study collected patient data from the First Affiliated Hospital of Nanjing Medical University between May 2021 and March 2024. Inclusion criteria mandated that all patients had undergone both CMR and right heart catheterization (RHC). CMR data included short-axis (SAX), four-chamber (4CH) and flow sequences, while RHC confirmed patient classification into one of three groups: non-PH, pre-capillary PH, or post-capillary PH. Clinical case records were used to compile and organize patient histories. The baseline characteristics of the study population are summarized in Table 1.

**Table 1**
Baseline characteristics of the study population.

| Characteristic | Value |
| --- | --- |
| Age, years (mean) | 81-16(48) |
| Non_PH, n(%) | 60(29%) |
| Pre-capillary PH, n(%) | 112(54%) |
| Post-capillary PH, n(%) | 32(15%) |
| Male, n(%) | 124(60%) |
| Ischemic heart disease(IHD), n(%) | 14(7%) |
| Dilated cardiomyopathy(DCM), n(%) | 60(29%) |
| Valvular heart disease(VHD), n(%) | 92(45%) |
| Chronic obstructive pulmonary disease(COPD), n(%) | 6(3%) |
| Rheumatoid immune disease(RID), n(%) | 68(33%) |
| Hyperthyroidism, n(%) | 4(2%) |
| Renal insufficiency, n(%) | 10(5%) |

*2.2. Right heart catheterization (RHC)*

RHC was performed using a 7.5F Swan-Ganz catheter (Edwards Lifesciences Corporation, CA, USA) via the internal jugular vein. The catheter position was confirmed using pressure waveforms or fluoroscopy when necessary. The manometer was zeroed at the level of the fourth intercostal space on the anterior chest wall, aligned with the midpoint of the bed surface. Measured parameters included right atrial pressure (RAP), right ventricular pressure (RVP), pulmonary arterial pressure (PAP), pulmonary artery wedge pressure (PAWP), cardiac output (CO), and pulmonary vascular resistance (PVR). PAWP was obtained by inflating the catheter balloon in a branch of the pulmonary artery. Cardiac output was measured using the thermodilution technique or the Fick method in the presence of shunts.

Pre-capillary PH was defined as a PAWP ≤ 15 mmHg, while post-capillary PH was defined as a PAWP > 15 mmHg with a mean pulmonary arterial pressure (mPAP) ≥ 20 mmHg, as measured by RHC at rest.

*2.3. Cardiac magnetic resonance (CMR) protocol*

All participants underwent CMR examinations using 3.0T scanners (Magnetom Skyra/Vida, Siemens Healthcare, Erlangen, Germany) during sinus rhythm with retrospective ECG gating and an 8-channel cardiac coil array. Velocity-Encoded 2D Phase-Contrast MR (PCMR) scans were acquired for the main pulmonary artery (MPA) with the imaging plane positioned 1 cm above the pulmonary valve. Imaging parameters included: spatial resolution = 1.4 × 1.4 mm², field of view (FOV) = 360 × 270 mm², slice thickness = 5 mm, echo time (TE) = 2.7–2.96 ms, repetition time (TR) = 4.95–5.25 ms, flip angle = 20°, bandwidth = 454 Hz/pixel, and velocity encoding (venc) = 130–150 cm/s. Scans were checked for velocity aliasing immediately after acquisition and repeated with adjusted venc if necessary. Cine images were acquired using a steady-state free precession (SSFP) sequence in four-chamber views and sequential short-axis slices from the atrioventricular ring to the left ventricular apex. Typical imaging parameters included: TR/TE = 3.24–3.44 ms / 1-2 ms, flip angle = 54°, and slice thickness = 8 mm.

## 3. Methods

### 3.1. Data preprocessing

Unified spatial resolution: The pixel spacing was standardized to 1.4 using a cubic spline interpolation algorithm. This value was chosen because the average pixel spacing of 204 samples was 1.4. The interpolation algorithm was optimized to minimize image distortion caused by extreme values (maximum or minimum), selecting a value close to the average, 1.4, as the standard. This approach ensures consistent spatial resolution across different images while retaining as much original image information as possible.

ROI segmentation: Using SAX data as an example, Unet++ [22] was employed to segment the cardiac region in short-axis (SAX) images, removing irrelevant data and providing high-quality input for the subsequent classification model [23]. Initially, a portion of the sample ROIs were manually labeled to train a basic segmentation model. The model was then used to automatically segment the remaining data, with non-compliant segmentation results manually corrected. After multiple iterations of training, segmentation, checking, and adjustment, all data were processed. Similar methods have been widely applied in medical image studies [24-26], and the processing of 4CH data follows the same workflow (see Figure 1). For SAX data, spatial resolution is standardized in Step 1. Step 2 involves iterative semi-automatic segmentation. Step 3 entails cropping the original images based on the binary segmentation results.

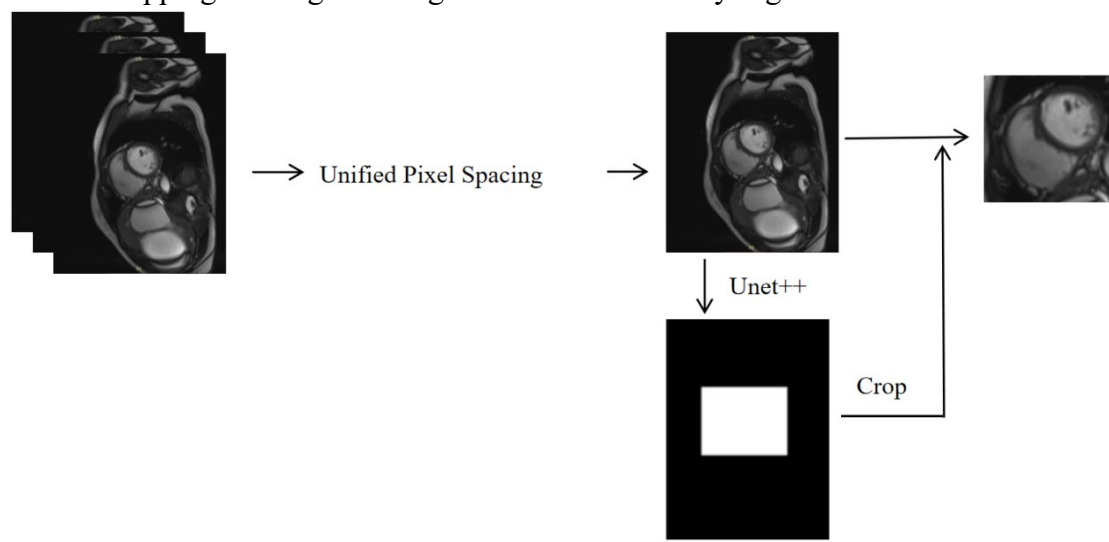

Figure 1. Data preprocessing workflow. For SAX data: Step 1 standardizes the spatial resolution; Step 2 performs iterative semi-automatic segmentation; Step 3 crops the original images based on the binary segmentation results.

Unified size: The SAX data were adjusted to [144, 144, 12, 5], where 144 represents the length and width, 12 represents the height, and 5 represents the time frames. Areas with length, width, or height smaller than [144, 144, 12] were zero-padded. The original SAX data consisted of 25 time frames, and five frames were selected with a step size of 5 as input to the model. Similarly, the 4CH data were uniformly adjusted to [160, 160, 5] to ensure consistency in input dimensions.

Data partitioning: Due to the imbalance in sample categories, with only 32 samples of post-capillary pulmonary hypertension, the Bootstrap method for data resampling was used. A total of 18 samples (approximately 10%) were selected for the test set, with six samples from each of the three categories. The remaining 186 samples were assigned to the training set. This process was repeated 35 times, with the final evaluation result being the average of the 35 experiments.

Clinical data preprocessing: Age normalization was performed using the Max-Min normalization method. The medical history, including IHD, DCM, VHD, COPD, portal hypertension, RID, hyperthyroidism, renal insufficiency, and gender, was previously categorized as binary (0 or 1), eliminating the need for further processing.

Pulmonary artery cross-sectional relative area change parameter (RAC): Swift et al. [27] experimentally demonstrated that PH patients exhibit reduced pulmonary artery elasticity due to increased pulmonary artery pressure. This is reflected by a decrease in the relative area change index of the pulmonary artery cross-section during the cardiac cycle. The calculation of RAC is provided in Equation (1).

$$RAC = \frac{MAX - MIN}{MIN} \qquad (1)$$

The cross-sectional area of the pulmonary artery can be obtained over the entire cardiac cycle in the flow sequence, as shown in Figure 2. *MAX* represents the maximum cross-sectional area within a cardiac cycle, and *MIN* represents the minimum cross-sectional area within the same cycle. This paper uses a U-Net model combined with an LSTM module to facilitate pulmonary artery cross-section segmentation. After calculating the area of each frame in the cardiac cycle using Pixel Spacing and pixel count, RAC is computed using Equation (1).

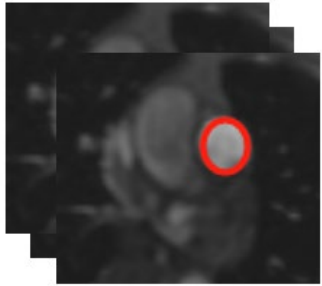

Figure 2. Illustration of pulmonary artery cross-section image segmentation. The image is a 3D representation combining the anatomy of the main pulmonary artery cross-section with the time

dimension, spanning a full cardiac cycle. The main pulmonary artery cross-section is shown within the red circle in the image. Segmentation of this region provides the MAX and MIN values in Equation (1), which are used to calculate the RAC.

*3.2. PH classification*

The framework of the proposed pulmonary hypertension classification network is illustrated in Figure 3. The workflow comprises three sequential stages: (1) acquisition of patients' CMR images, followed by automated segmentation of the pulmonary artery cross-section for relative area change RAC quantification; (2) pre-processing of short-axis and four-chamber (4CH) view images through resolution standardization and region-of-interest (ROI) alignment, enabling the DC-Transformer module to extract spatiotemporal features; (3) deployment of a graph convolutional network (GCN) to model inherent associations between imaging biomarkers and clinical parameters. Subsequent sections will present comprehensive technical specifications for each architectural component.

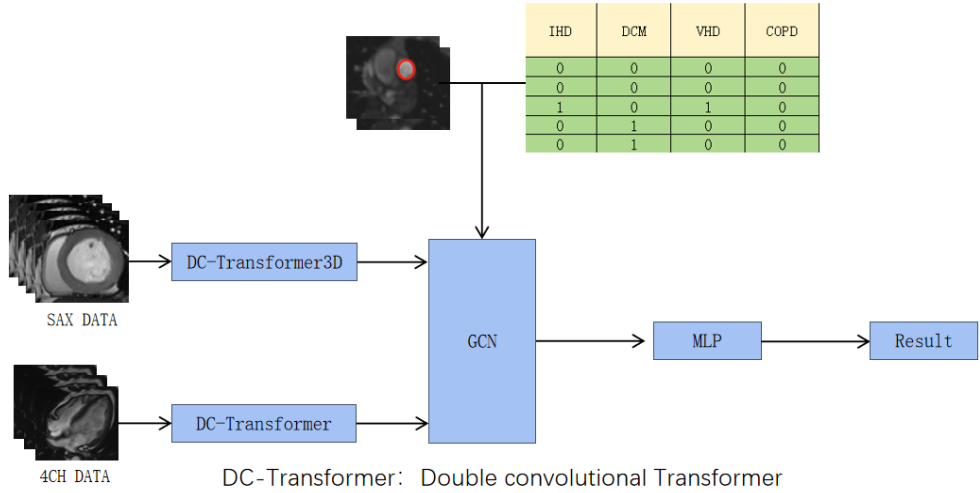

Figure 3. Architecture of the Multimodal PH Classification Network. The figure illustrates the integration of pulmonary artery cross-sectional images with clinical tabular data. After calculating RAC from the images, the data are integrated in parallel with clinical information into the GCN for analysis. The input and output of the GCN are treated as feature vectors, and the MPL classifier directly generates the classification results.

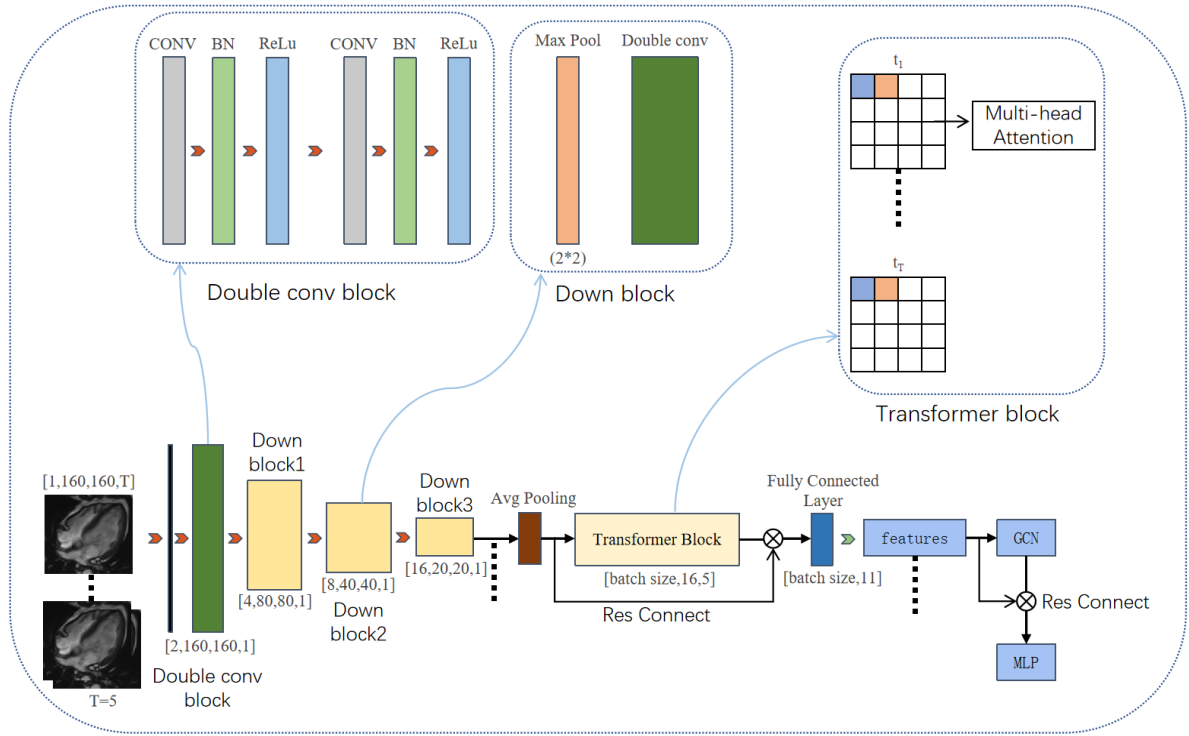

Figure 4. Structure of the DC-Transformer. Using 4CH data as an example, with 5 time frames (T=5), each time frame first passes through the DC and downsampling modules to extract deep spatial features. Next, time features are extracted by concatenating the frames in temporal order, followed by the application of the multi-head self-attention mechanism to each time frame.

### 3.2.1. DC-transformer

As shown in Figure 4, the network architecture first extracts spatial features using two convolutional layers followed by batch normalization (BN) layers. Each convolutional layer is followed by a ReLU activation function, enhancing the model's representational capacity and improving training efficiency. The batch normalization layer normalizes the output of each layer, alleviating internal covariate shift and accelerating training, thereby improving the model's generalization. The Transformer module plays a key role in extracting temporal features from image sequences. TThe self-attention mechanism allows the Transformer to capture long-term dependencies across different time steps in the sequence, effectively capturing dynamic changes in the time series. Compared to traditional recurrent neural networks (RNNs), the self-attention mechanism in the Transformer offers superior parallelism and more efficient sequence data processing.

The input video frames or consecutive image sequences first undergo two convolutional and downsampling operations to extract the spatial features of each frame. These spatial features offer a deep understanding of the spatial structure within a single image. However, extracting spatial features alone is insufficient for capturing temporal relationships within image sequences. Therefore, the spatial features of each frame are concatenated to form a multi-frame temporal feature representation. The concatenated features retain the spatial information of individual frames and form a temporal feature flow, preparing for further temporal feature extraction. The Transformer uses a self-attention mechanism that facilitates the weighted aggregation of each frame in the input temporal features. This process enables the extraction of dynamic temporal features that represent the spatial features of each frame in the image sequence. Additionally, it integrates information along the temporal dimension, facilitating the model's understanding of temporal dependencies between frames.

*3.2.2. Graph convolutional network*

Graph convolutional network (GCN) are a type of neural network designed for graph-structured data [28 29]. The core idea of GCN is to perform convolution operations using information from the graph structure, including nodes and edges. In this approach, both the input and output of the graph neural network can be interpreted as feature vectors, with the propagation process detailed in Equation (2).

$$H^{(l+1)} = \sigma(\hat{A} H^{(l)} W^{(l)}) \qquad (2)$$

Here, $H^{(l+1)}$ represents the node representation matrix at layer *l+1*, $H^{(0)}$ is the feature matrix of the input graph. $W^{(l)}$ is the learnable weight matrix at layer *l*. σ is the activation function. $\hat{A}$ is the normalized adjacency matrix, defined as shown in Equation (3).

$$\hat{A} = D^{-1/2}(A + I)D^{-1/2} \qquad (3)$$

Here, *A* is the adjacency matrix of the graph, *I* is the identity matrix, *D* is the degree matrix.

In multimodal learning, features derived from different modalities have distinct semantics and representations. Traditional methods primarily process these features in isolation, using individual network modules for each modality without feature fusion. However, to capture interactions between modality features, feature fusion is essential. In the approach, feature extraction is first performed for each modality using separate networks, followed by fusion with structured clinical data (e.g., age, gender, medical history). Additionally, RAC is included as a clinical input.

To effectively manage the interactions between multimodal and clinical features, it is essential to construct the graph adjacency matrix. This adjacency matrix represents both the connection relationships between nodes and the similarity or correlation between each pair of nodes. Graph relationships are encoded in the adjacency matrix, which governs how information propagates and aggregates between nodes. This forms the basis for the model to learn graph structural information. Each element in the adjacency matrix indicates the presence of an edge between two nodes. The novel graph relationship construction method proposed in this paper is illustrated in Figure 5.

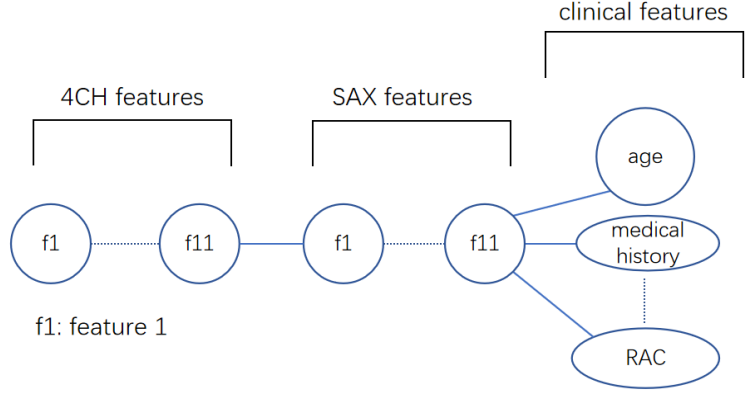

Figure 5. Construction of the graph adjacency matrix using graph relationships. The term "features" in the figure refers to 11 nodes, each with individual feature values. The 4CH&SAX features are obtained by extracting image features using the previously discussed DC-Transformer. There are a total of 11 nodes in clinical features, which are encoded through clinical information. RAC is calculated using Equation (1).

In the propagation process of graph neural networks, an excessive number of nodes in one modality can lead GCN to prioritize information from that modality, potentially neglecting information from others. To address this, we equalize the number of image feature nodes with that of clinical information nodes using a fully connected layer. Finally, we demonstrate through the graph node ablation experiment that selecting 11 nodes is optimal.

In the initial experiment, we attempted to construct a graph relationship with three nodes, using features from three modalities, where each node was represented as an 11-dimensional vector. However, the results were suboptimal, even worse than those obtained using only the MLP. After a detailed analysis of the GCN propagation process, we found that, in this graph relationship, features from the same modality are constrained by linear relationships, as shown in Figure 6. In contrast, the relationship between feature values and prediction outcomes is more complex and higher-dimensional. Therefore, we split the feature values across different nodes, as illustrated in Figure 7. Finally, we demonstrate through GCN ablation experiments that the fusion of GCN is advantageous.

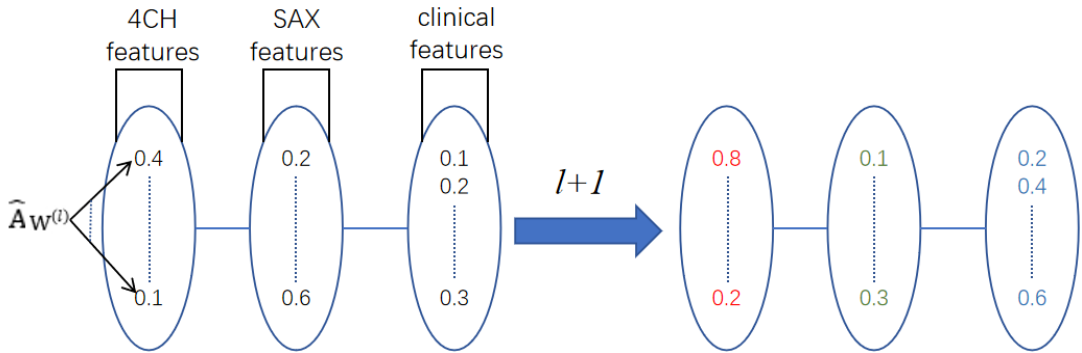

Figure 6. GCN Propagation Process. The colored numbers in the figure represent a single propagation step of the GCN, where numbers of the same color are subject to identical linear constraints. Specific weight values are selected to visually illustrate the linear constraints applied to f1 through f11.

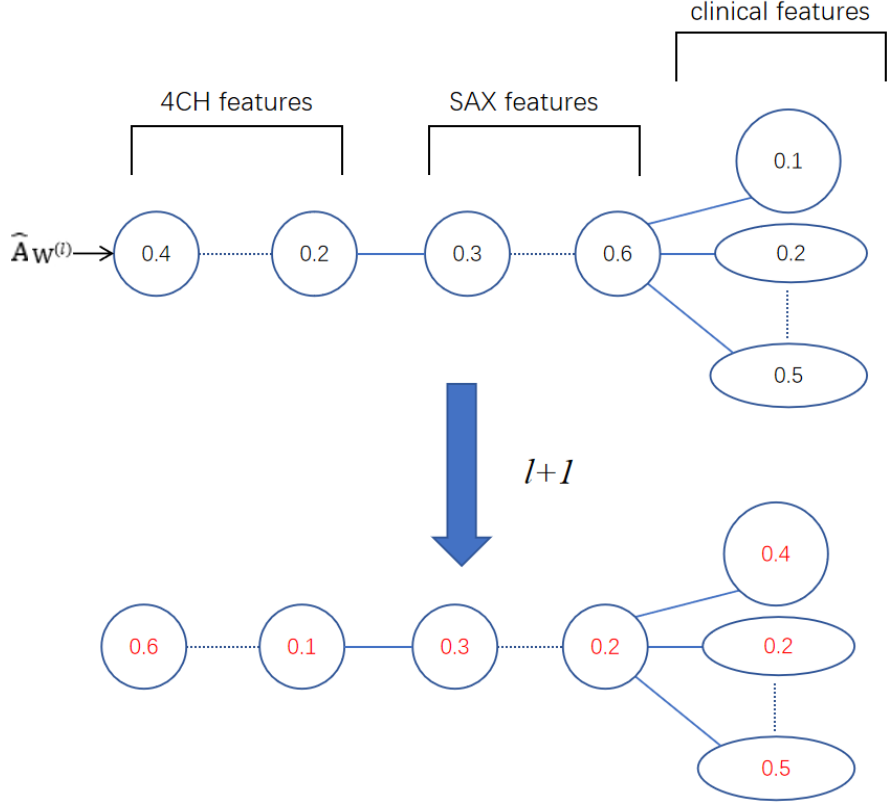

Figure 7. After Splitting the Nodes, GCN Can Express Higher-Dimensional Relationships. The red numbers in the figure simulate a single propagation step of the GCN, where the linear constraints are broken.

### 3.3. Loss function

The model has been trained with the cross-entropy loss function, as shown in Equation (4). This function is one of the most frequently employed loss functions in machine learning and deep learning, especially in classification problems. It quantifies the difference between the predicted probability distribution derived from the model and the actual probability distribution of the labels.

$$L = -\sum_{i=1}^{C} y_i \log(p_i) \tag{4}$$

Here, $C$ is the number of sample classes. $y_i$ is the ground truth label of the sample in class $y$. $p_i$ is the probability predicted by the model that the sample belongs to class $i$. $L$ is the loss value.

### 3.4. Training parameter settings

The model is trained using a Tesla V100 GPU with 32GB of GPU memory, the PyTorch 2.3.1 platform, and Python version 3.9. During the training phase, the Adam optimizer is used, with a cross-entropy loss function and a learning rate set to 0.006. The model's batch size is set to 2, and the number of training epochs is set to 30.

### 3.5. Evaluation

The primary evaluation metrics include three components: average accuracy (ACC), the area under the receiver operating characteristic curve (AUC), and class-wise AUC. The class-wise AUC is calculated by treating samples from one class as positive and the remaining classes as negative, followed by computing the AUC for that class.

## 4. Results and Analysis

### 4.1. Results

As shown in Table 2, performance was evaluated across 35 random splits of the training and testing datasets. The model achieved an average accuracy (ACC) of 0.71 and an area under the curve (AUC) of 0.81. The highest observed performance was an ACC of 0.88 and an AUC of 0.92, while the lowest performance was an ACC of 0.61 and an AUC of 0.68.

Initial experiments revealed that the choice of dataset splits had a significant impact on the model's performance. After thorough analysis, we attributed this variability to the limited number of collected samples. To further investigate the impact of sample size on model performance, we conducted experiments with incremental training samples.

**Table 2.**

The average performance of the model on the test set. The values are expressed as mean ± standard deviation across the 35 experiments.

|  |  | ACC | AUC | Sensitivity | Specificity |
| --- | --- | --- | --- | --- | --- |
| Non PH | Average | $0.785_{\pm 0.072}$ | $0.745_{\pm 0.118}$ | $0.666_{\pm 0.207}$ | $0.845_{\pm 0.107}$ |
| Pre-capillary PH | of | $0.846_{\pm 0.067}$ | $0.863_{\pm 0.068}$ | $0.828_{\pm 0.161}$ | $0.854_{\pm 0.080}$ |
| Post-capillary PH | 35 | $0.838_{\pm 0.065}$ | $0.834_{\pm 0.109}$ | $0.709_{\pm 0.179}$ | $0.902_{\pm 0.064}$ |
| ALL | experiments | $0.734_{\pm 0.060}$ | $0.814_{\pm 0.063}$ | —— | —— |

### 4.1.1. Influence of the Number of Training Samples

In this experiment, the test set remains constant, while the number of training samples is incrementally increased from 71 to 186, with a step size of 5. Training samples were randomly selected at each step. The results, illustrated in Figure 8, show the model's performance improvements as the sample size increases.

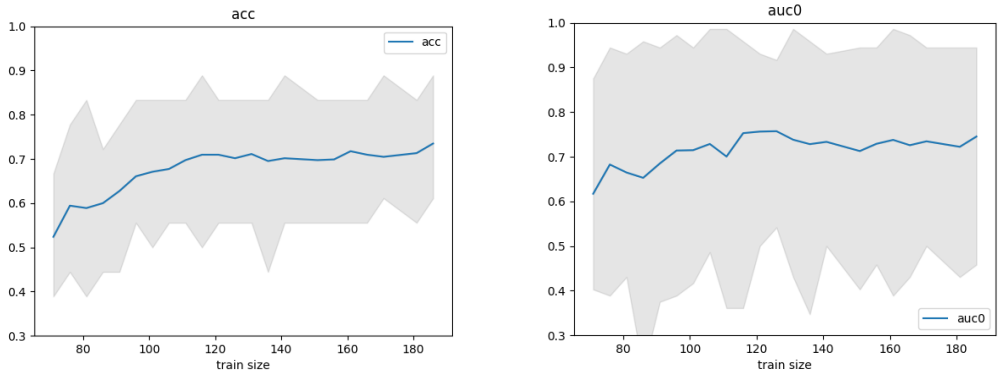

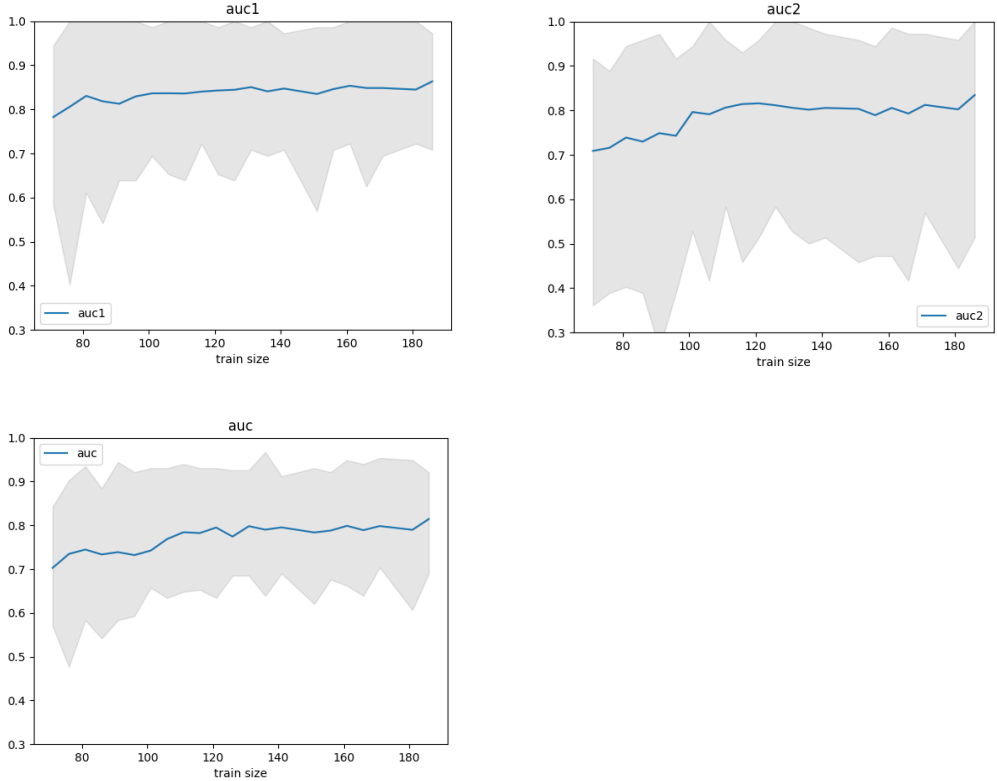

Figure 8. Performance of the incremental training sample experiment. The horizontal axis represents the training sample size, increasing from left to right, and the vertical axis represents the corresponding metric values. The gray area represents the range between the maximum and minimum values across 35 experiments. ACC: accuracy; AUC: area under the receiver operating characteristic curve; AUC0: non-PH AUC; AUC1: AUC of pre-capillary PH; AUC2: AUC of post-capillary PH.

### 4.1.2. Influence of the GCN Module

To assess the contribution of the GCN to the classification model, we conducted an ablation experiment. In this setup, multimodal features were directly input into a Multilayer perceptron (MLP) classifier, while all other training parameters remained constant.

Across 35 random sampling experiments, the inclusion of GCN improved the model's performance in several key metrics. Specifically, the model using GCN showed an improvement in ACC from 0.70 to 0.73, the highest AUC increased from 0.916 to 0.921, and the lowest AUC rose from 0.68 to 0.689. The detailed results of the ablation experiment are presented in Table 3.

**Table 3.**
Results of the GCN ablation experiments. The values are expressed as mean ± standard deviation across the 35 experiments. The experiment ensured that the training parameters and dataset partitioning were kept completely consistent. AUC0: Non-PH AUC; AUC1: AUC of pre-capillary PH; AUC2: AUC of post-capillary PH.

|  | AUC | ACC | AUC0 | AUC1 | AUC2 |
|---|---|---|---|---|---|
| Full_model | **0.814**$_{\pm 0.063}$ | **0.734**$_{\pm 0.060}$ | **0.745**$_{\pm 0.118}$ | **0.863**$_{\pm 0.068}$ | **0.834**$_{\pm 0.109}$ |
| No_GCN | 0.804$_{\pm 0.068}$ | 0.695$_{\pm 0.066}$ | 0.742$_{\pm 0.101}$ | 0.861$_{\pm 0.066}$ | 0.809$_{\pm 0.112}$ |

*4.1.3. Influence of the Number of Graph Nodes*

To demonstrate the derivation of using 11 nodes in the aforementioned graph neural network, we scaled the image features extracted by the DC-Transformer to different feature dimensions using a trainable fully connected layer, while keeping other experimental conditions unchanged, to vary the number of nodes in the graph. The experimental results, shown in Table 4, indicate that the best performance is achieved with 11 nodes, further supporting our hypothesis.

**Table 4.**
Results of the Graph node ablation experiments. The values are expressed as mean ± standard deviation across the 35 experiments, and the training/test partition follows the full model partition method described in this paper. AUC0: Non-PH AUC; AUC1: AUC of pre-capillary PH; AUC2: AUC of post-capillary PH.

| nodes | AUC | ACC | AUC0 | AUC1 | AUC2 |
|---|---|---|---|---|---|
| 5 | 0.797$_{\pm 0.067}$ | 0.711$_{\pm 0.074}$ | 0.727$_{\pm 0.122}$ | 0.847$_{\pm 0.080}$ | 0.817$_{\pm 0.106}$ |
| 9 | 0.807$_{\pm 0.062}$ | 0.723$_{\pm 0.072}$ | 0.766$_{\pm 0.112}$ | 0.857$_{\pm 0.069}$ | 0.789$_{\pm 0.114}$ |
| 10 | 0.811$_{\pm 0.058}$ | 0.723$_{\pm 0.082}$ | **0.776**$_{\pm 0.105}$ | 0.863$_{\pm 0.079}$ | 0.794$_{\pm 0.107}$ |
| **11** | **0.814**$_{\pm 0.063}$ | **0.734**$_{\pm 0.060}$ | 0.745$_{\pm 0.118}$ | **0.863**$_{\pm 0.068}$ | **0.834**$_{\pm 0.109}$ |
| 12 | 0.810$_{\pm 0.051}$ | 0.723$_{\pm 0.060}$ | 0.774$_{\pm 0.092}$ | 0.856$_{\pm 0.075}$ | 0.800$_{\pm 0.112}$ |
| 13 | 0.808$_{\pm 0.069}$ | 0.719$_{\pm 0.067}$ | 0.771$_{\pm 0.098}$ | 0.861$_{\pm 0.073}$ | 0.792$_{\pm 0.118}$ |
| 22 | 0.803$_{\pm 0.055}$ | 0.709$_{\pm 0.069}$ | 0.761$_{\pm 0.108}$ | 0.842$_{\pm 0.075}$ | 0.804$_{\pm 0.106}$ |
| 44 | 0.789$_{\pm 0.073}$ | 0.707$_{\pm 0.066}$ | 0.733$_{\pm 0.119}$ | 0.837$_{\pm 0.082}$ | 0.796$_{\pm 0.102}$ |

## 5. DISCUSSION

Pulmonary hypertension (PH) is a rare, progressive cardiovascular disease characterized by elevated pulmonary artery pressure, ultimately leading to right heart failure and poor patient prognosis [30-33]. Accurate identification of PH is crucial for improving patient outcomes, and the precise differentiation between pre-capillary and post-capillary PH types is essential for guiding treatment strategies [34-36]. In clinical practice, a model that predicts only the presence of PH without classifying its subtypes is insufficient. Our model addresses this gap by predicting not only the presence of PH but also differentiating between pre-capillary and post-capillary PH types. This dual capability provides clinicians with actionable insights, enhancing informed decision-making and potentially reducing diagnostic workload [33].

*5.1. Clinical Applicability*

The ability to classify PH subtypes is especially valuable in clinical settings. Pre-capillary PH, often associated with conditions such as primary pulmonary hypertension (PH), requires targeted therapies such as endothelin receptor antagonists, phosphodiesterase 5 inhibitors, guanylate cyclase stimulators, prostacyclin analogues, and prostacyclin receptor agonists. In contrast, post-capillary PH, typically associated with left heart disease, requires management of underlying cardiac conditions [37 38]. The structural and functional parameters of the left and right ventricles and pulmonary artery can be assessed using cardiac CMR. By automating this classification process, our model can streamline diagnostic workflows, reduce reliance on invasive procedures such as RHC, and improve patient care. Moreover, the integration of CMR data provides a non-invasive alternative to RHC, which, despite being the gold standard, carries inherent risks and may not be universally accessible.

Our model stands out due to its comprehensive approach to PH diagnosis and classification. Previous studies have made significant advances in PH prediction but fall short in subtype differentiation. For example, Liu et al. [39] developed an AI model using electrocardiogram data to predict PH versus non-PH, achieving an AUC of 0.88. Similarly, Kusunose et al. [21] developed a deep learning model based on chest X-rays (CXR) to detect exercise-induced pulmonary arterial hypertension, achieving an AUC of 0.74. While these models demonstrate commendable performance, they are unable to classify PH subtypes, limiting their clinical utility.

In contrast, Hirata et al. [20] used logistic regression to classify non-PH, pre-capillary PH, and post-capillary PH using echocardiogram data, achieving AUCs of 0.789, 0.766, and 0.742, respectively. However, conventional machine learning methods, such as logistic regression, require significant time and effort for feature selection [40]. Our deep learning model overcomes this limitation by automating feature extraction, achieving an AUC of 0.81 and significantly reducing the time and effort required for model development. This represents a significant advancement in both efficiency and performance.

*5.2. Multimodal Feature Fusion with GCN*

Feature fusion remains a challenging task. In this study, we innovatively propose a multimodal feature fusion framework using graph convolutional networks (GCN) and experimentally validate its effectiveness in integrating imaging features with clinical features. As demonstrated in Table 4, the fusion performance reaches optimal levels when the graph node count corresponds to the number of clinical features. However, severe node-quantity mismatch leads to significant performance degradation, with results underperforming even compared to non-GCN baseline methods. This observation highlights that node-clinical feature alignment is critical for harnessing GCN's full potential. Notably, while the methodology was developed for pulmonary hypertension (PH) diagnosis, its extensibility to other deep learning-based disease diagnostic systems shows substantial promise.

*5.3. Influence of the Number of Training Samples*

We have thoroughly explored the relationship between training sample size and model performance, as shown in Figure 8. The results show that when the sample size reaches approximately 120, the growth rate of AUC and ACC slows, but continues to increase. This suggests that 120 samples can support model stability. However, increasing the sample size may further enhance model performance.

The model demonstrates better stability in recognizing category 1. This may be attributed to class imbalance in the dataset, where the model learns the majority class (e.g., category 1) more effectively. Conversely, this suggests that adding more data, particularly for minority classes, may improve the model's stability. Additional samples expose the model to diverse features and patterns, improving its learning and generalization, thereby enhancing stability and accuracy in recognizing different categories. Therefore, continuous collection and refinement of data samples are necessary to improve the model's performance and applicability in real-world scenarios.

*5.4. Limitations*

Despite its strengths, our study has several limitations. First, the "black-box" nature of deep learning models presents challenges in interpretability, which is crucial for gaining clinician trust and facilitating adoption. Second, the dataset used to train our model is imbalanced, which may affect its generalizability.

*5.5. Future work*

Future studies should aim to incorporate larger and more diverse datasets to overcome this limitation. Furthermore, our research focuses on specific imaging modalities. Future research could explore multi-step decision methods, where initial imaging results are used to determine whether additional diagnostic tests are necessary. This approach could optimize resource utilization and reduce the burden on patients.

## 6. CONCLUSION

Our deep learning-based diagnostic method for non-PH, pre-capillary PH, and post-capillary PH is efficient and stable, offering substantial potential to alleviate the burden on patients. Furthermore, the technological innovations presented in this approach are highly adaptable and can be extended to other diagnostic fields, demonstrating broader applicability and potential for future advancements.

**CRediT authorship contribution statement**

**Fubao Zhu**: Writing - review & editing, Supervision, Project administration, Funding acquisition. **Yang Zhang**: Writing - original draft, Software ,Validation, Methodology, Data curation, Formal analysis. **Gengmin Liang**: Data curation, Resources, Investigation. **Jiaofen Nan**: Writing - review & editing, Funding acquisition. **Yanting Li**: Writing - review & editing, Funding acquisition, **Chuang Han**: Writing - review & editing, Funding acquisition. **Danyang Sun**: Writing - review & editing, Validation. **ZhiguoWang**: Funding acquisition. **Chen Zhao**: Writing - review & editing, Validation, Formal analysis, **Wenxuan Zhou**: Data curation, Investigation. **Jian He**: Resources, Data curation. **Yi Xu**: Writing - review & editing, Resources. **Iokfai Cheang**: Resources, Data curation, Writing - review & editing. **Xu Zhu**: Resources, Data curation. **Yanli Zhou**: Writing - review & editing, Investigation, Conceptualization, Resources. **Weihua Zhou**: Conceptualization, project administration.

**Acknowledgments**

This study received support from the National Natural Science Foundation of China (Grant Numbers: 62476255, 62106233, 62303427, and 82370513), the Science and Technology Innovation Talent Project of Henan Province University (Grant Number: 25HASTIT028), The Zhongyuan Science and Technology Innovation Outstanding Young Talents Program, the Henan Science and Technology Development Plan (Grant Number: 232102210010, 232102210062).

**Declaration of competing interest**

The authors of this manuscript declare no relationships with any companies whose products or services may be related to the subject matter of the article.

**Ethical Approval**

Institutional review board approval was obtained (Ethics Approval Number: 2022-SR-506).

**Informed Consent**

All participants have signed the informed consent form.